\documentclass[a4paper,11pt,a4paper,onecolumn]{article}
\pdfoutput=1 

\usepackage[utf8]{inputenc}
\usepackage[T1]{fontenc}
\usepackage[greek,american]{babel}

\usepackage{fullpage}
\usepackage{rotating}

\usepackage{booktabs} 
\usepackage{tabularx}

\usepackage[charter]{mathdesign}
\usepackage{microtype}

\usepackage{textcomp}

\usepackage{comment}

\usepackage{graphicx, color}

\usepackage{hyperref}
\hypersetup{colorlinks=false,
		urlcolor=blue,
		citecolor=red,
		pdfauthor={Stefan Wagner}}

\usepackage{amsmath}

\usepackage[small,bf]{caption} 
\makeatletter
\g@addto@macro\bfseries{\boldmath}
\makeatother

\usepackage{siunitx}

\newcommand{\um}{\SI{}{\micro\metre}}

\newcommand{\lm}{$\lambda_\mathrm{max}$}

\renewcommand{\emph}[1]{\textbf{#1}}

\title{A hybrid organic/inorgaic scintillator for high performance measurements}
\author{Stefan Wagner$^1$\footnote{Corresponding author: stefan.wagner@apc.in2p3.fr}, Marco Grassi, Anatael Cabrera}
\date{\today}

\begin{document}

\maketitle
\begin{abstract}
\noindent New experimental research programs in the field of neutrino physics are calling for new detectors with large masses, high energy resolution and good background rejection capabilities. This paper presents a novel hybrid organic/inorganic scintillator, which is able to improve on all three aspects simultaneously. This scintillator consists of microscopic grains of inorganic crystals suspended in an organic scintillating carrier medium. Due to multiple scattering off the crystals, this scintillator appears opaque over longer distances and is intended for use in specialized detectors. Thanks to the crystal phase it can natively incorporate a large variety of elements in large quantities, so that a sufficiently large detector can reach elemental loadings on the ton- or multiton scale. At the same time, this composition can produce very high light outputs and provides additional particle identification capabilities. This scintillator concept is expected to provide significant advantages for future neutrino experiments, like searches of neutrinoless double beta experiments and reactor antineutrino physics.
\end{abstract}


\tableofcontents


\section{Introduction}
The recent years have seen tremendous advancements in the experimental investigation of (anti)neu\-tri\-nos and their properties. The precision measurement of the previously unknown leptonic mixing angle $\theta_{13}$ by the Daya~Bay, Double Chooz and RENO experiments is perhaps the most notable example~\cite{DB, DC, RENO}. All three experiments employ large-volume organic liquid scintillators for antineutrino detection to achieve a sufficiently large target mass. In a similar manner, large scintillation detectors are used with great success in the search for neutrinoless double beta decay. The best limit to date was achieved by the KamLAND Zen experiment with about 380~kg of isotopically enriched xenon in a kiloton liquid scintillator detector~\cite{KamLANDZen}. To reach a much larger active mass, the upcoming SNO+ experiment will also make use of a massive liquid scintillator volume of 780~t and will be able to load 1.3~t of $^{130}$Te into the detector~\cite{SNOplus}. Likewise, in the field of neutrino oscillations the next-generation experiment JUNO will use 20~kt of liquid scintillator~\cite{JUNO}. But the ever increasing detector sizes present major challenges to experimentalists and enegineers, not to speak of the project costs. Experiments like Aurora or Candles, use scintillating crystals as detector materials instead~\cite{Aurora, Candles}. They provide better energy resolution and already incorporate the target elements in the crystal structure, but this comes at the expense of the achievable total detector mass.

This paper presents a novel hybrid scintillator, which combines the advantages of both inorganic crystals and organic liquids, and can provide high light yields, high elemental loading and good particle identification capabilities at the same time. The concept of this microcrystal scintillator is presented in the following. Section~\ref{two} introduces the general working mechanism and the mechanical design requirements are presented in section~\ref{threefour}. The paper then concentrates on the expected performance of this new scintillator in section~\ref{five}. Special focus is given to the large selection of potential materials, the very high elemental loadings, as well as to the superior light yield and high energy resolution achievable with such a configuration. The additional pulse shape discrimination capabilities are also discussed. Section~\ref{six} concludes this paper with an outlook on future work related to this scintillator concept.

\section{Working principle}\label{two}
The microcrystal scintillator is composed of microscopic grains of inorganic scintillating crystals suspended in an organic liquid scintillator. Over large distances, the heterogeneous mixture will appear opaque due to strong multiple scattering, but at short distances from the point of creation, light can be detected or extracted, for example by means of wavelength shifting fibers. This kind of scintillator is designed for use in detectors in which the light is detected close to the point of creation. This includes small, handheld detectors, as well as large highly segmented detectors~\cite{SoLiD}, and the recently proposed Liquido detector concept~\cite{Liquido,Wurm}. 
	\begin{figure*}[t]
	\begin{centering}
	\includegraphics[width=0.5\textwidth]{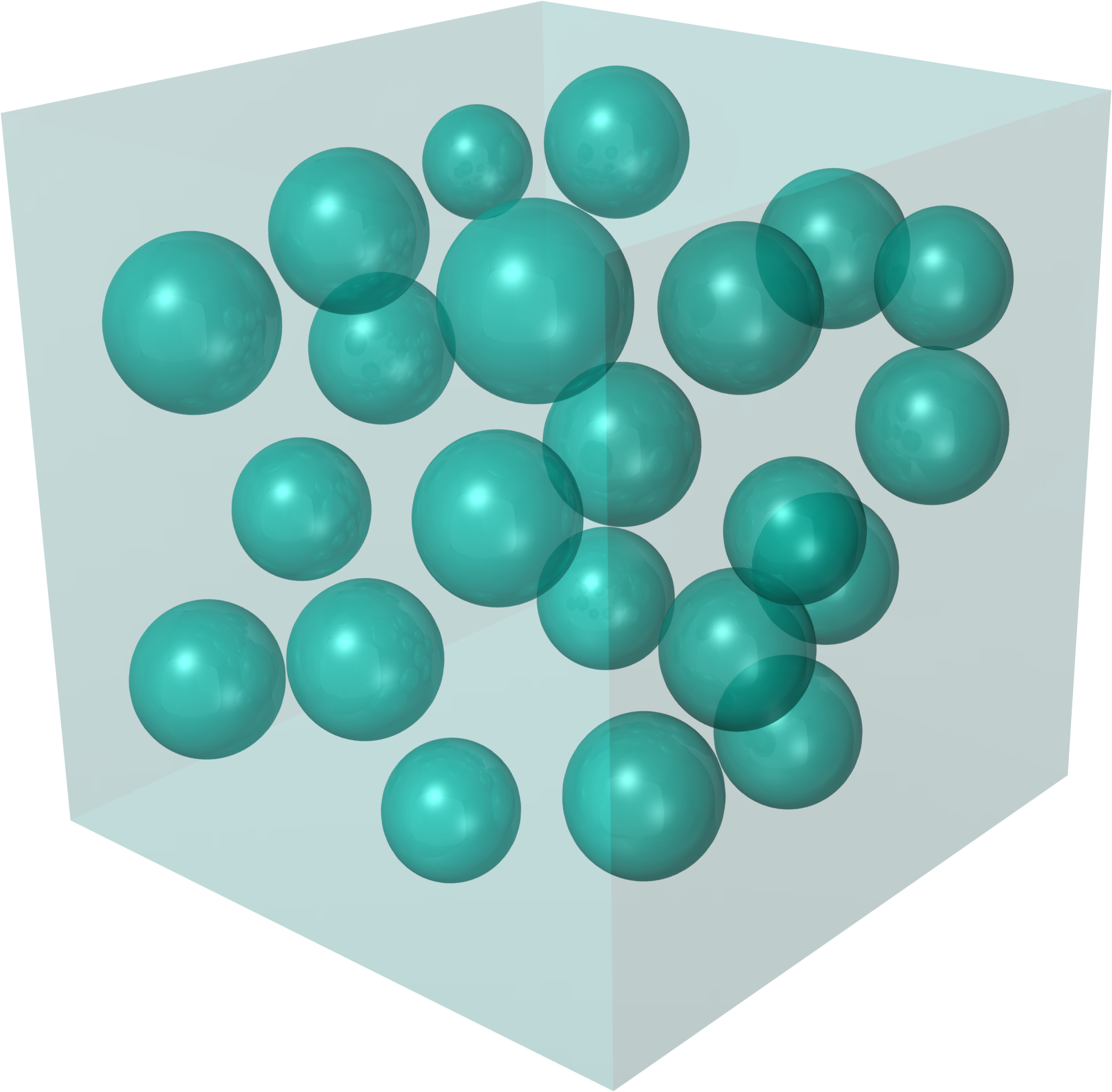}
	\caption{\small Rendered illustration of the microcrystal scintillator composition. Small grains of inorganic scintillating crystals are suspendend in a scintillating carrier material. The crystals are held in suspension either in form of a stabilized colloidal suspension or via a gel-like medium.\label{GrainBox}}
	\end{centering}
	\end{figure*}
The latter is based on a fiber lattice in a highly scattering scintillating medium. In this approach multiple scattering of the emitted scintillation photons confines the light to a small region around the event vertex. The scintillator appears opaque over longer distances, since photons cannot directly reach photodetectors on the outside of the volume. The photons are instead collected by an array of wavelength shifting fibers in the detection volume and observed by a photon detector at the end of the fiber. The confinement leads to a high light collection efficiency on the fibers and a very high vertex and pattern resolution of events in the detector. Different particles can be identified by their energy deposition patterns, leading to a significant background suppression. Alpha particles and low energy electrons will show a single interaction site, while gammas typically deposit their energy in multiple Compton scattering events. Positrons show a very distinct topological signature created by their initial energy deposition and two gamma quanta from its subsequent annihilation, and can therefore be well distinguished from other types of events.
	\begin{figure*}[t]
	\begin{centering}
	\includegraphics[width=0.75\textwidth]{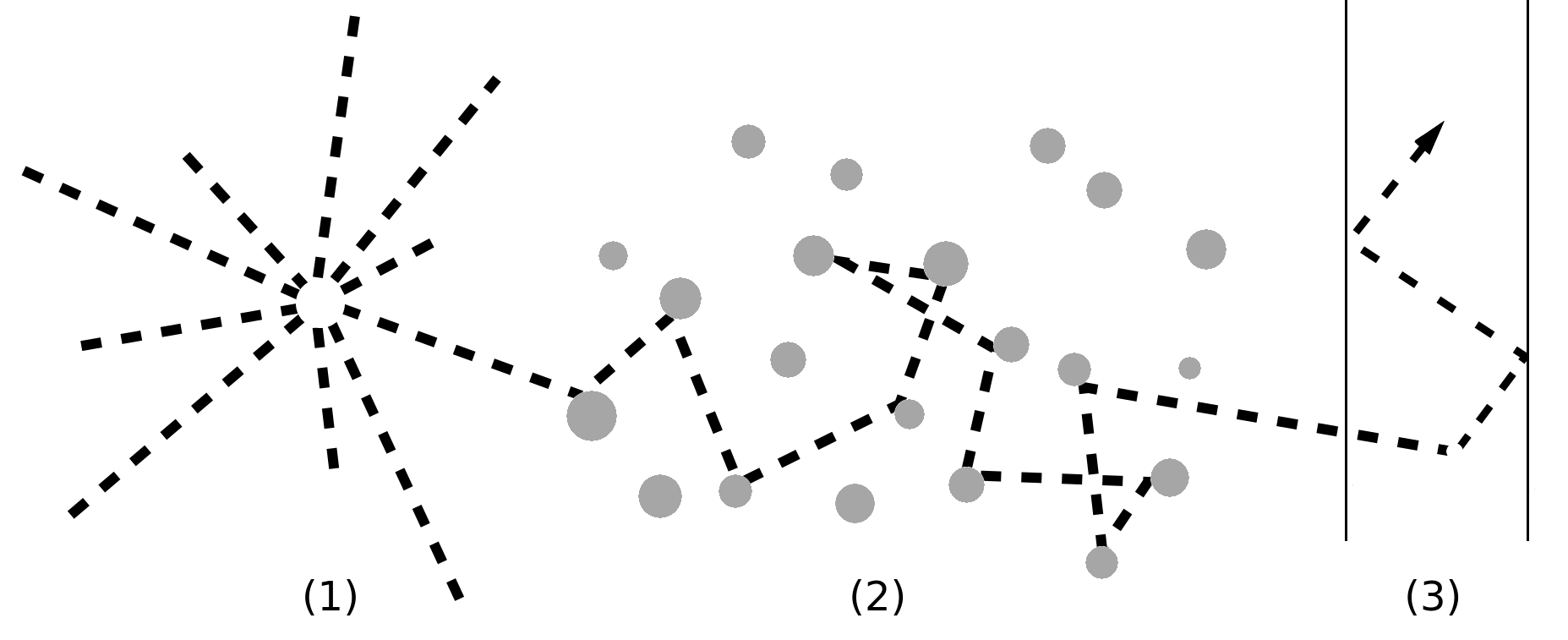}
	\caption{\small Working principle of the microcrystal scintillator. 1) An ionizing particle excites the scintillator and creates scintillation light. 2) The emitted light is scattered off the crystal grains multiple times, essentially leading to a random walk in the medium. This limits the effective travel distance of the photon and confines the light to a region around the origin. 3) The photon hits a fiber, where it can be extracted from the medium and propagated to a photodetector.\label{Principle}}
	\end{centering}
	\end{figure*}\\
\\
The microcrystal scintillator concept is illustrated in figure~\ref{GrainBox}. The crystal grains serve three important roles at the same time: they are the primary generators of scintillation light with high light yield, act as scattering centers for the emitted photons, and supply the target mass. The carrier medium fixes the particles in their places, creates additional light with fast timing, and shifts the primary scintillation spectra to the desired wavelengths. When a charged particle propagates through the microcrystal scintillator, it excites both the crystal and the liquid phase. The crystal grains act as the primary scintillation sources and produce a large number of photons per MeV of deposited energy. The organic phase adds additional scintillation light. The emitted optical photons then propagate through the materials and are scattered off the crystal grains. The photons undergo multiple scattering processes in quick succession and essentially perform a random walk in the medium. They diffuse into a cloud and are spatially confined to a small region around the event vertex, until they either reach a wavelength shifting fiber or are absorbed. This process is illustrated in figure~\ref{Principle}.


\section{Microcrystal scintillator design\label{threefour}}
The mechanical design of a microcrystal scintillator depends decisively on the grain sizes. The microscopic grains of crystal scintillator are the core constituent of this formulation. The particles must be small enough to be able to remain in suspension, and their sizes should also be much smaller than typical particle tracks in the mixture, in order to achieve a uniform response of the scintillator. The required grain dimensions are between 0.1 and 10~{\um}.




Two options are available to create the grains with the desired dimensions. In a top-down approach larger grains of crystal material can be brought to the required sizes with adequate crushing and milling techniques. Jet milling, for example, can readily produce crystalline powders with average particle sizes in the range of 1 to 10~{\um} and some cases the grain sizes can even be brought down to about 0.1~{\um}. Milling is routinely performed on an industrial scale and large batches can be processed by specialized service providers at low costs. Without further processing, however, the resulting grains are likely of irregular shapes and sizes.

In a bottom-up approach, microcrystals can be grown from solution. This is an alternative option for some materials and can produce very small, spherical and evenly-sized grains. Sizes of 0.1~{\um} and below are achievable with this method. Scintillating microcrystals that are commonly produced by controlled precipitation from solution include ZnS, CdS and CaF$_2$.~\cite{Trindade, Dalas, Markovic}.  If available, this option is preferrable, as it produces grains of higher quality, but it may have a lower throughput.\\ 
\\
The particle dimensions then fix the properties of the carrier medium. Particles in of size range between 1 and 10~{\um} will slowly but steadily settle in a traditional liquid scintillator. In this size range the particles are subject to Stokes' flow. The terminal velocity of a grain is reached when the gravitational force is in equilibrium with the Stokes drag and buoyancy forces. To achieve sufficient stability of such a mixture over the lifetime of the experiment, the carrier medium is required to have high dynamic viscosity. In a medium with a dynamic viscosity of 10000~cP the settling speed can be reduced to under 10$^{-10}$~m/s, which corresponds to about 3~mm/year, which is probably tolerable in most practical situations. A medium of 10000~cP viscosity has a honey-like consistency and is still flowable. The advantage of such a system is that the mixture remains fluid and can be easily poured into a detection volume and will completely fill arbitrary shapes or detector volumes with delicate structures. 

The concentration of particles in the suspension also plays a role. Above a certain value the the relative viscosity of the mixture, i.e. the viscosity of the mixture compared to the pure fluid, increases with increasing grain concentration. This effect leads to mutual blocking of the particles and also hinders their sedimentation. Above a certain threshold concentration it might cause a transition of the suspension from a fluid to an elastic solid, which might be desired or not. Another possibility to stabilize the system is to directly add thickening agents to the medium and allow solidification in place.

At particle sizes around 1~{\um} and below one enters the colloidal regime. In a colloidal suspension the particles are small enough to be subject to molecular effects. While the relative influence of gravity decreases, Brownian motion tends to maintain the grains in suspension. On the other hand, attractive van der Waals forces between different grains can lead to aggregation into larger clusters, if they are not balanced by other effects. The clusters may then become large enough to precipitate from the mixture. This phenomenon is well studied and the creation of stable colloidal suspensions is a common task in chemical engineering~\cite{Kronberg, Mewis, Adair}. Different methods are available to stabilize a colloidal suspension. One way is by means of steric stabilization. The grains are covered in suitable polymers, which form a soft shell around the hard crystal grains and inhibit aggregation. The overlap of these shells lead to a repulsive effect when two particles approach each other, which stabilizes the suspension.

The rheological behavior of a colloidal suspension is complex and depends on the material of the crystals, the suspending fluid, the concentration and shape of the particles, as well as on the additives used to stabilize the system. If the concentration of suspended particles is low, a stabilized colloidal suspension can be rather fluid and more similar to traditional scintillator. High particle concentrations can lead to jamming of the grains and lead to the transition into an elastic solid when a critical volume fraction of roughly~0.5 is reached~\cite{Faraday,Poon}. If the suspension should still be flowable, the maximum volume fraction of the crystal phase should be well below this value.\\
\\
The crystals can be chosen from a wide range of highly performant materials. The main boundary condition for the selection of the crystal material is the chemical compatibility of the crystal with the medium, i.e. that they do not degrade or dissolve. In particular, many crystal scintillators are hygroscopic and deteriorate when they are exposed to humidity. For this reason the medium should in general be free of water. Possible materials are viscous hydrocarbon oils or polysiloxanes. While the former must be handled in a controlled atmosphere to avoid absorption of water vapors, the latter tend to expel water by themselves.

The organic phase can be used to tune the scintillation spectra with help of fluorescent molecules and can even be scintillating themselves. This is a preferred option to maximize the light yield of the configuration and can be realized by mixing aromatic components into the bulk fluid. For a hydrocarbon-based carrier, this could be linear alkyl benzenes, while for a silicone-based bulk medium it is possible to use phenylated polysiloxanes. In combination with secondary fluors they have been demonstrated to work as efficient scintillators with high light yield~\cite{Quaranta}. Secondary fluors in the carrier shift the primary scintillation photons (both from the crystal and the organic phase) to higher wavelengths, away from the absorption spectra of the crystals and the carrier, and towards the sensitive range of the fibers. Ideally, the emission spectrum of the fluors has a large overlap with the absorption spectrum of the fibers to maximize the light collection efficiency. For example, PPO or PTP can be dissolved in the carrier at a concentration of several~g/l. The fluor concentration can be lower than in common liquid scintillator mixtures, since multiple scattering increases the probability to encounter a fluorophore within a given distance.


\section{Properties and performance \label{Performance}\label{five}}

\subsection{Scattering}
	\begin{figure*}[t]
	\begin{centering}
	\includegraphics[width=0.5\textwidth]{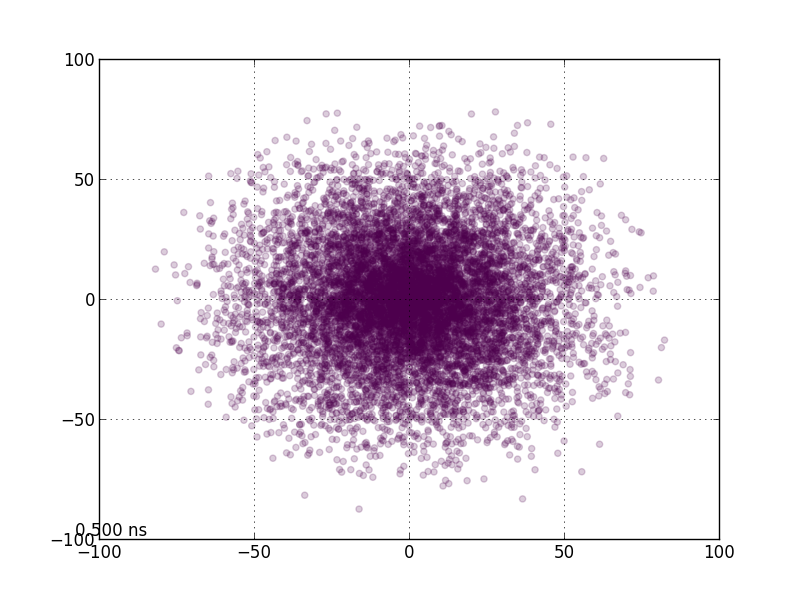}\includegraphics[width=0.5\textwidth]{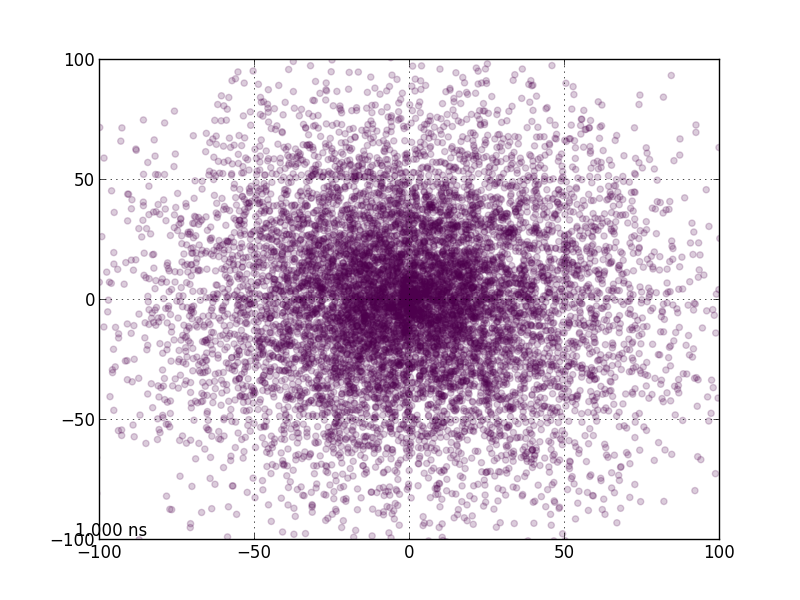}\\
	\includegraphics[width=0.5\textwidth]{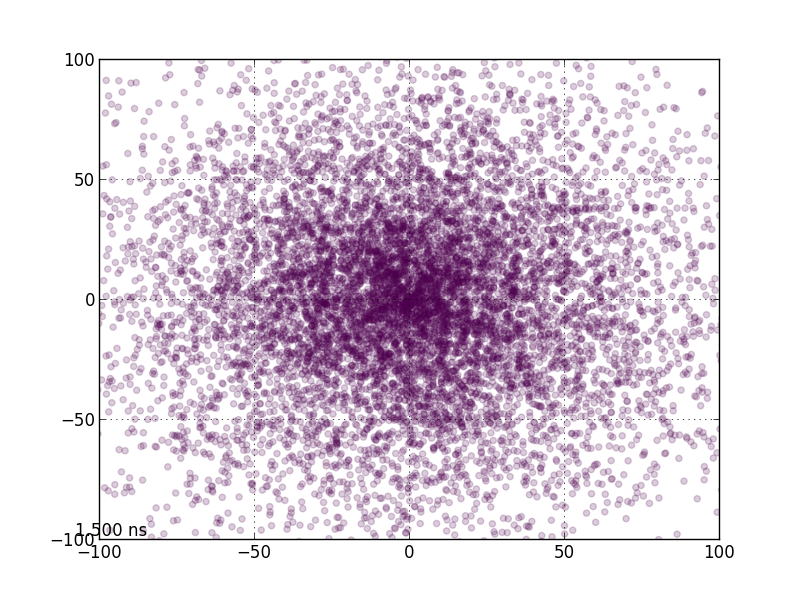}\includegraphics[width=0.5\textwidth]{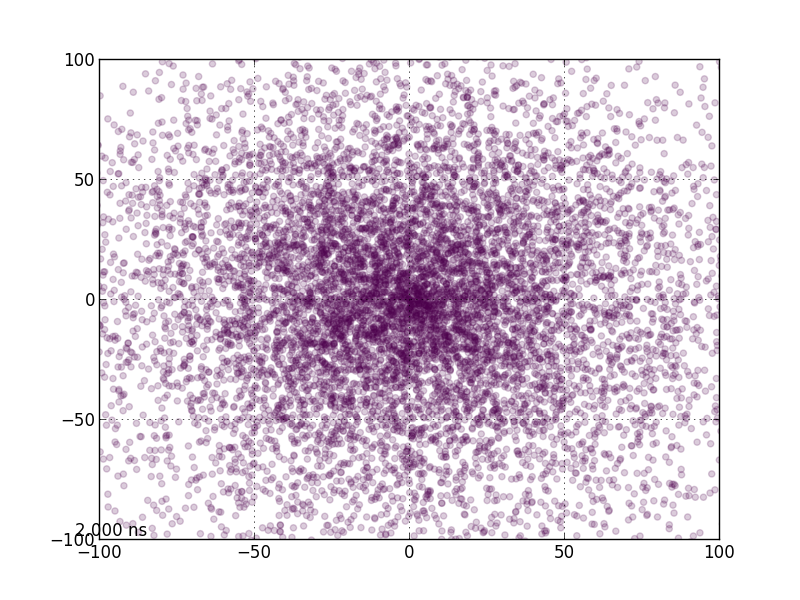}
	\caption{\small Simulated diffusion of 10000 scintillation photons in a microcrystal environment. The photons are injected at the center in positive x-direction. A random homogeneous distribution of crystal grains of 8.0~{\um} and equal volumetric proportions of liquid and crystal phase was assumed. The photon wavelength is 400~nm and the refractive indices of crystals and the medium are 1.44 and 1.40 respectively. In this configuration the mean free path between two scatter processes is 3.5~{\um}. The panels show the photon positions after 0.5~ns, 1.0~ns, 1.5~ns and 2.0~ns. The axis scales are in mm. The photons are confined to a volume around the point of creation and a majority remains within about 5~cm around the point of creation.\label{Scattering}}
	\end{centering}
	\end{figure*}
The behavior of optical scattering within the medium depends critically on the size of the particles and the wavelength of the photons. For particle sizes in the range between 0.1 to 10~{\um} the scattering of visible light is governed by Mie scattering~\cite{Mie}. Mie scattering is a complex phenomenon and depends on a number of parameters, but the cross section increases with particle sizes and the difference of the refractive indices between particle and medium, and decreases with the photon wavelength. Mie scattering shows a strong angular dependency and favors scattering angles close to 0{\textdegree} and 180{\textdegree} with respect to the incident photon direction. The influence of backscattering decreases with increasing particle size, so that Mie scattering becomes very forward-pointed.


To obtain an estimation of the scattering behavior in the microcrystal environment, the propagation of photons under multiple Mie scattering was simulated with a custom Monte Carlo code. The scattering angles and cross sections were calculated with help of the \textit{miepython} package, which was developed and validated for the simulation of light propagation in turbid media~\cite{OMLC}. The photons were propagated in steps. After each step the travel distance until the next scattering process and the scattering angle were sampled. Figure~\ref{Scattering} shows the light confinement due to multiple Mie scattering in a microcrystal environment. All photons were injected at the origin in positive x-direction. Even though each scattering process has a strong preference for forward-pointed angles, the vast number of interactions leads to an effective random walk and an isotropic diffusion of the photons. A photon can propagate several meters in the material before being absorbed, but the effective travel distance is limited to several centimeters. The scattering length can be tuned by varying the crystal sizes and concentrations, thus adjusting the size of the photon cloud.

The Mie scattering probability is strongly influenced by the difference of the refractive indices of the two phases. By variation of the refractive index of the organic phase it is possible to tune scattering behavior. If the refractive indices of both phases are carefully matched, scattering becomes minimal and the scintillator becomes translucent over longer distances. The translucency can be further enhanced if the grain dimensions and concentrations are lowered. At a point where the mixture becomes sufficiently transparent it can also be used more traditional detector concepts, while still providing high light yield, high elemental loading, and good particle discrimination capabilities. Such a transparent configuration of inorganic crystals in a gel scintillator has been successfully produced and studied for dark matter experiments~\cite{Spooner,Tovey}.\\
\\
The more isotropic Rayleigh scattering is much weaker, but it may have some influence in the optical behavior if there is a high amount of large molecules in the organic phase. This could either be long polymer chains in a viscous carrier or other additives like surfactants and stabilizers, or large metal-organic complexes for increased loading.


\subsection{Materials and loading}

\begin{table*}[t]
\centering
\begin{tabular}{@{}lrllll@{}}
Scintillator		& LY	[MeV$^{-1}$]	& \lm~[nm]	& $\tau$~[ns]		& $\beta\beta$	& \\
\midrule
CdWO$_4$		& 27000			& 495		& 12000			& Cd	 		& \cite{Moszynski}\\
NaI:Tl			& 38000			& 415		& 250			& ---			& \cite{StGobain}\\
CaF$_2$:Eu		& 19000			& 435		& 940			& Ca			& \cite{StGobain}\\
CsI:Tl			& 54000			& 550		& 1000			& ---			& \cite{StGobain}\\
ZnSe:Te			& 55000			& 645		& 50000			& Zn,Se,Te	& \cite{Schotanus}\\
SrI$_2$:Eu		& 120000			& 435		& 1200			& ---			& \cite{Vanloef}\\
CaI$_2$			& 86000			& 550		& 550			& Ca			& \cite{Hofstadter}\\
CaMoO$_4$		& 400			& 520		& 17				& Ca, Mo		& \cite{Belogurov}\\
BaF$_2$:Ce 		& 1300			& 220		& <1				& Ba			& \cite{Melcher}\\
LaBr$_3$:Ce		& 63000			& 380		& 16				& Ce			& \cite{StGobain}\\
LaCl$_3$:Ce		& 61000			& 356, 387	& 30				& Ce			& \cite{Vanloef2002}\\
\bottomrule
\end{tabular}
\caption{Properties of some selected inorganic crystal scintillators. The table gives an overview over the  variety of elements which can be incorporated in the crystal phase. The information on light yield, emission wavelength and scintillation decay is only indicative and can depend on the experimental conditions, particularly on the dopant concentration. Column four shows constituent elements which are candidates for double beta decay.  \label{ScintillationData}}
\end{table*}

There is a large variety of crystal scintillator materials and in principle any crystal can be chosen for a microcrystal scintillator formulation, provided that it is chemically compatible with the carrier. In this way there is a large selection of elements that can be natively introduced into the detector in large amounts. Table~\ref{ScintillationData} shows several crystal scintillators and their particular constituent elements, as well as their scintillation properties. Many elements have important properties for specific physics cases. For example, elements like gadolinium or lithium have high neutron capture cross sections and are frequently loaded into liquid scintillators for neutron detection. High-Z elements like lead or bismuth are well suited for detection of X-rays and gamma radiation.

Other elements offer particularly low thresholds for neutrino or antineutrino detection. In a dedicated survey it was found that chlorine and bromine are among the most promising candidates for the detection of antineutrinos with energies below the inverse beta decay threshold of 1.8~MeV. The effective energy threshold for charged-current antineutrino interaction on $^{35}$Cl and $^{79}$Br are at 1.189 and 1.268~MeV respectively. Together with their comparably high natural abundance and interaction cross section they are probably the most suitable elements for charged-current detection of low energy antineutrinos. It turns out that LaBr$_3$:Ce and LaCl$_3$:Ce, which contain a large fraction of these elements in their crystal lattice, are very bright scintillators with excellent timing properties (see Table~\ref{ScintillationData}). With a microcrystal formulation based on these materials very high target masses can be realized. Assuming equal volume fractions of crystal and organic phase, 1~m$^3$ of microcrystal scintillator can hold roughly 2.54~t of LaBr$_3$:Ce, providing $6.1\cdot10^{27}$ targets of $^{79}$Br (0.81~t). The total crystal mass of LaCl$_3$:Ce is smaller due to its lower density, but it provides a larger number of targets: $1.1\cdot10^{28}$ targets of $^{35}$Cl (0.64~t). A large detector based on microcrystalline LaBr$_3$:Ce or LaCl$_3$:Ce could find applications in the the study of the low energy part of the geoneutrino energy spectrum or of the reactors antineutrino spectra. Both crystals are widely used due to their high energy resolution and are produced on an industrial scale (e.g.~\cite{SaintGobain}).

Another important physics case are searches for neutrinoless double beta decays. Table~\ref{ScintillationData} also lists the double beta isotopes present in several crystal scintillators.
As an example, even a compact 1~m$^3$ volume of a microcrystal scintillator based on CdS with a volume fraction of 10\% contains 375.5~kg of natural cadmium and 28.2~kg ($1.5\cdot10^{26}$ atoms) of the double beta decaying isotope $^{116}$Cd. This would already surpass the target masses of current experiments by nearly two orders of magnitude~\cite{Aurora}. 
It shall be emphasized here, that the active mass in the microcrystal approach is obtained without any isotopic enrichment, which grants a serious cost advantage\footnote{Isotopic enrichment, though expensive, would yield another order of magnitude of target mass.}. If activated with appropriate dopants (like indium or tellurium), CdS can be very fast and have high light yields~\cite{BNL}.




It is currently not possible to accomplish similar levels of loading with traditional liquid scintillators. Significant development effort is required to design stable and soluble metal-organic complexes and special care has to be taken during production and handling of the scintillator. Nevertheless, the dopant concentrations are usually limited to a low percentage of the scintillator mass before either the optical properties or the stability of the mixture degrade.\\
\\
Other candidate isotopes for neutrinoless double beta decay can also be employed. Elements which can be easily introduced via the bulk crystal phase include calcium, selenium, and tellurium. Other elements which are not encountered in the crystal lattice can be incorporated as dopants. Crystal scintillators are routinely activated with foreign elements to enhance the light output and/or to optimize the emission spectra and decay times. Cerium contains potential double beta decay isotopes. $^{136}$Ce is believed to undergo double positron emission and hence particularly well suited for study in a detector with good topological resolution, since it can exploit the unique pattern of two positrons. In addition, $^{142}$Ce is a potential double beta decaying isotope. Cerium is used as a dopant in the aforementioned lanthanum halide scintillators, but a larger amount of cerium can be obtained with CeCl$_3$ (46000 photons, 25ns)~\cite{Vanloef2001b} or CeBr$_3$ (68000 photons, 17ns)~\cite{Shah}.

Loading more exotic elements into a detector also becomes feasible in a microcrystal environment. Indium- and neodymium-loaded liquid scintillators have been proposed for solar neutrino physics and double beta decay experiments, respectively~\cite{Raghavan,Chen}. While these elements do not typically make up the crystal lattice, they can be introduced as dopants into the crystals at levels of a few percent. 

In cases in which a single crystal species cannot fulfill all desired requirements, it is also possible to introduce grains of two different types of crystals in order to profit from their respective advantages. An example would be to use one crystal with high light yield and another one which contains isotopes with high neutron capture cross section, as it is done with mixed LiF/ZnS scintillator screens for neutron detection.

\subsection{Light yield and energy resolution}
Crystal scintillators typically have much higher light yields than organic liquid scintillators (see Table~\ref{ScintillationData}). For instance, NaI:Tl creates about 45000~photons/MeV, which surpasses the light output of organic liquid scintillators significantly. Some more recently developed crystals, like BaBrI:Eu, LaBr$_3$:Ce or SrI$_2$:Eu, can have light outputs of around $10^5$~photons/MeV, which is about 10 times larger than of typical organic liquid scintillators. A high primary light yield is critical to minimize the statistic contribution to the energy resolution. If the intrinsic non-statistic contribution is not too large, a relative energy resolution of \textless~2~\%$_\sigma$ at 1~MeV seems within reach. Single crystals of LaBr$_3$:Ce were reported to have a total intrinsic energy resolution of 2.9~\%$_\mathrm{FWHM}$ at 622~keV~\cite{Vanloef2002}, which translates to about 1~\%$_\sigma$ energy resolution at 1~MeV, assuming a sufficiently linear behavior. Such a high energy resolution would be exceptional in a large-scale scintillation detector. In comparison, the next-generation JUNO experiment aims for a yet unprecedented energy resolution of 3~\%$_\sigma$~\cite{JUNO}. A high energy resolution is also a crucial condition for neutrinoless double beta decay searches in order to resolve the signal from the irreducible two-neutrino double-beta decay background. A high light output is also important when looking for low-energy signals, like nuclear recoils or X-rays. The microcrystals could still provide several thousand primary photons in these cases.

In principle, the heterogeneous microscopic structure of the scintillator and the different light yields of the organic and inorganic phase give rise to two different energy scales. A particle of low kinetic energy may deposit most of its energy in the phase in which it is created and the mean number of photons depends on whether the particle originates from the crystal or the medium. However, it was shown by simulation that the response of the microcrystal system already becomes uniform at electron energies as low as 50~keV, which is well below the typical energy region of interest. Above this limit electrons have sufficient energy to leave the original phase and traverse both phases. The light yield and scintillation characteristics are then determined by both phases. Protons, alphas and other heavy particles, on the other hand, have much shorter tracks than electrons and their threshold for a uniform response lies significantly higher. This can be used for particle discrimination, as it will be discussed in the next paragraph.

The heterogeneous structure leads to another interesting phenomenon. Since particles of sufficiently high energy deposit their energy in both phases, the total light yield of the microcrystal scintillator is somewhat lower than that of pure crystals. But due to the higher stopping power of crystals, particles tend to deposit more energy in the crystal phase than in the surrounding medium. The combined scintillation characteristics are thus still dominated by the crystal phase. For example, a Geant4 simulation of the effective total light yield of an exemplary microcrystal system with a volume fraction of 0.5 and light yields of 10000~photons/MeV in the organic and 40000~photons/MeV in the organic phase showed an effective light yield of about 36000~photons/MeV. There might be transfer of excitation energy from the carrier to the crystal grains, which is not included in the simulation and may increase the light yield even further. This also means that the amount of the more expensive crystal material can be reduced in favor of the less expensive organic carrier, while maintaining the favorable performance of the crystal scintillator to a large extent.

\subsection{Timing and pulse shape discrimination}
While crystal scintillators have excellent light yields, their fluorescence decay times are often much longer than those of organic liquid scintillators. The primary decay times typically range from several hundres to several thousand nanoseconds. 
A slow response can be partly compensated by the fast timing of the scintillating organic phase. The combined signal of the microcrystal scintillator is a superposition of a fast and a slow component and the sharp peak of the liquid phase can be used to obtain precise timing information of the event. The delayed photons from the crystal scintillation are then collected over time to provide an accurate energy information.

Some crystals, however, have ultrashort decay times that surpass those of organic liquids. The semiconductor crystals CdS and ZnO, for instance, show ultrafast timing with appropriate dopants~\cite{Lehmann}. BaF$_2$ is another scintillator with sub-nanosecond decay times~\cite{Laval}. Such crystals could be used for an extremely fast timing. On the other hand, the above mentioned scintillators have only moderate light yields. But in situations in which timing is important, these crystals could be used together with grains of a slower but brighter crystals.\\
\\
The timing difference between the two phases also leads to additional pulse shape discrimination power, especially at low particle energies where the heterogeneity of the formulation comes increasingly into play. When a charged particle does not possess enough energy to create extended tracks (e.g. recoil nuclei, alphas or low-energy electrons), its energy is deposited mostly in the phase from which it originates. The resulting scintillation waveform is consequently either more crystal-like or more liquid-like.

Electrons in this energy range can be created by Compton scattering of gammas. Gammas tend to deposit their energy in interactions with the \mbox{high-$Z$} elements present in the crystals, rather than in the \mbox{low-$Z$} liquid phase and the Compton electrons show a slightly more crystal-like scintillation wave form. In contrast, fast neutrons lose their energy primarily in collisions with hydrogen nuclei in the liquid phase, while they traverse the crystals nearly unhindred. The recoil protons have very short tracks and deposit their energy predominantly in the liquid phase, so their scintillation time profile is dominated by the liquid phase, producing a faster but weaker signal. 

The recoils of other nuclear recoils might also be distinguished thanks to the often noticable intrinsic pulse shape differences of crystal scintillators in response to different particle types. This possibility was already demonstrated in a similar environment with a transparent, index-matched mixture based on CaF$_2$. The observed pulse shape differences allowed to discriminate between calcium and fluor recoils~\cite{Spooner}.\\
\\
The different energy loss mechanisms of different particles not only leads to different pulse shapes, but also to changes in the emission spectra. This color information is usually not exploited in scintillation detectors. In the microcrystal scintillator, on the other hand, it leads to a different diffusion behavior of the photons, which can be observed as a different intensity distribution of the photon cloud around the origin. Since the Mie scattering cross section depends on the wavelength of the photon (both directly and indirectly via the wavelength-dependent refractive index of the medium), photons with longer wavelengths tend to scatter slightly less than photons with shorter wavelengths, so that they can diffuse farther away from the origin. Thus, an emission spectrum leaning towards longer wavelengths, as in the case of alpha particles, leads to a slightly larger photon cloud with a broader intensity distribution pattern, which might be exploited for particle discrimination.

\section{Conclusions and outlook\label{six}}
The microcrystal scintillator concept is a promising option for future neutrino experiments based on new detector types. It was shown that this approach is able to improve on all important aspects of scintillator detectors: it provides large masses, good energy resolution and background rejection capabilities. While there are many possible use cases for this type of scintillator, these advantages make it particularly attractive for future searches for neutrinoless double beta decay.

A microcrystal scintillator can also be produced in very large quantities and incorporate many tons of crystal material, while the fluidity of the mixture facilitates the construction of large-volume monolithic detectors. Experiments with active isotope masses of several tons thus seem reachable with this approach, which would mark an improvement of several orders of magnitude over current experiments. In addition, a microcrystal scintillator is expected to be significantly cheaper than a corresponding arrangement of single crystals, which is another practical advantage for the construction of large detectors.\\
\\
While the expected behavior and performance of microcrystal scintillator systems have been investigated with help of simulations, several aspects, including energy transfer mechanisms between the organic and inorganic phase, or the influence of local inhomogeneities on the scattering, cannot be reliably investigated in that manner. Other properties depend on the particular design of the system and cannot be treated in a general manner. These questions should be resolved experimentally. For this purpose, batches of different microcrystal scintillator configurations are to be produced at the APC, Paris. They will be experimentally studied under different aspects and for their potential use in the Liquido project.



\section{Acknowledgments}
We are thankful to C.~Buck and B.~Gramlich from the Max Planck Institut für Kernphysik (MPIK), Heidelberg, for fruitful discussions and useful comments, as well as K.~Thaler of Rmory Studios for the rendered illustration of the microcrystal scintillator concept.


\end{document}